\documentclass[11pt,a4paper]{article}
\usepackage{jcappub}
\usepackage{euscript,epsfig,amsmath,amssymb}
\usepackage{amsfonts,latexsym}

\newcommand{\be}[1]{\begin{equation}\label{#1}}
\newcommand{\ee}{\end{equation}}
\newcommand{\ba}[1]{\begin{eqnarray}\label{#1}}
\newcommand{\ea}{\end{eqnarray}}
\newcommand{\rf}[1]{(\ref{#1})}
\newcommand{\nn}{\nonumber}

\begin{document}

\title{Dark matter and dark energy from quark bag model}

\author{Maxim Brilenkov$^{1}$,}
\author{Maxim Eingorn$^{2}$,}
\author{Laszlo Jenkovszky$^{3}$}
\author{and Alexander Zhuk$^{4}$}

\affiliation{$^{1}$Department of Theoretical Physics, Odessa National University,\\ Dvoryanskaya st. 2, Odessa 65082, Ukraine\\}

\affiliation{$^{2}$Physics Department, North Carolina Central University,\\ Fayetteville st. 1801, Durham, North Carolina 27707, USA\\}

\affiliation{$^{3}$Bogolyubov Institute for Theoretical Physics, Kiev 03680, Ukraine\\}

\affiliation{$^{4}$Astronomical Observatory, Odessa National University,\\ Dvoryanskaya st. 2, Odessa 65082, Ukraine\\}

\emailAdd{maxim.brilenkov@gmail.com} \emailAdd{maxim.eingorn@gmail.com} \emailAdd{jenk@bitp.kiev.ua} \emailAdd{ai.zhuk2@gmail.com}

\abstract{We calculate the present expansion of our Universe endowed with relict colored objects -- quarks and gluons -- that survived hadronization either as
isolated islands of quark-gluon "nuggets"\, or spread uniformly in the Universe. In the first scenario, the QNs  can play the role of dark matter. In the
second scenario, we demonstrate that uniform colored objects can play the role of dark energy  providing the late-time accelerating expansion of the Universe.}

\maketitle

\flushbottom

\section{Introduction}

Over two decades ago \cite{LasloYad1,LasloYad2} (see also \cite{Tillmann}) the accelerated expansion of the early Universe was derived from a quark bag model
with the proper equations of state (EoS). It was called tepid \cite{LasloYad1,LasloYad2} or little \cite{Tillmann} inflation, in view of its moderate scales,
compared to the better known earlier inflation. However, occurring at a later time (when the temperature $\sim 200$ MeV) and smearing a lot of the earlier
effects, it may have important consequence for the observable Universe.

The derivation was based on a quark-gluon bag EoS completing the Friedmann equations. Our Universe was cooling down along the "hot"\, (i.e. quark-gluon) branch
of the EoS until it reached the point/area of the transition to the confined hadron phase. In most of the papers along these lines (see \cite{LasloYad1} and
references therein) the process of inflation terminates by a phase transition (hadronization) to the state of colorless objects. The details of the phase
transition, depending on unknown confining forces, are poorly known and leave much room for speculations.

In the present paper, we consider the possibility that a small fraction of colored objects -- quarks and gluons -- escaped hadronization. They may survive as
islands of colored particles, called quark-gluon nuggets (for brevity sometimes also quark nuggets (QNs)). This possibility was first considered by E. Witten
\cite{Witten} and scrutinized further in \cite{Applegate, Farhi, Chandra}. In his paper \cite{Witten}, E. Witten discusses the possibility that QNs
can survive even at zero temperature and pressure. If so, the "hot"\, quark-gluon phase in the form of QNs may affect the present expansion of the
Universe. Indeed, our investigation shows that nuggets can contribute to dark matter provided that their interaction with ordinary matter is weak.

The size distribution of QNs was calculated in \cite{Bhatta,Bhatta2}. The authors found that a large number of stable QNs exists in the present Universe. They
also claimed that QNs could be a viable candidate for cosmological dark matter. The survival probability of these QNs, i.e. the question whether the primordial
QNs can be stable on a cosmological time scale, is a key issue, and it was studied by a number of our predecessors. In particular, the authors of \cite{BASR},
using the chromoelectric flux tube model, have demonstrated that the QNs will survive against baryon evaporation if the baryon number of the quark matter
inside the nuggets is larger than $10^{42}$ which is a rather conservative estimate. A scenario where the Universe would be closed with QNs with the baryon
number density window $10^{39\div40}\leq N\leq 10^{49}$ or, in other words, the proverbial cosmological dark matter, containing $90\%$ or more of all matter in
the Universe, is made of QNs, was considered in the paper \cite{ARS}. The special role of the strange quark matter in the phase transition, both in the context
of the early Universe and in compact stars, was discussed in \cite{Gh}. A relativistic model for strange quark stars was proposed in \cite{Kalam} (see also
\cite{Zhitnitsky} for a different approach to get compact quark objects). Quark matter is believed to exist at the center of neutron stars \cite{Perez}, in
strange stars \cite{Drake} and as small pieces of strange matter \cite{Madsen}. The latter can result in ultra-high energy cosmic rays \cite{Madsen2,Madsen3}.
The search (in lunar soil and with an Earth orbiting magnetic spectrometer) for cosmic ray strangelets may be the most direct way of testing the stable strange
matter hypothesis.

Another possibility is that a very small (to be specified!) fraction of colored objects -- quarks and gluons -- survived after the phase transition in the form
of a perfect fluid  uniformly spread within the colorless hadronic medium. This picture is physically less  motivated than the nugget model. We suppose that
this fluid has the same thermodynamical properties as the quark-gluon plasma (QGP). Therefore, we call it as a QGP-like perfect fluid. Nevertheless, it is of
interest to investigate cosmological consequences of such assumption. In our paper, we demonstrate that such fluid can provide an alternative (with respect to
the cosmological constant) explanation to the late-time accelerating expansion of the Universe. It is worth mentioning that quark-gluon plasma as dark matter
in the halo of galaxies was investigated in the paper \cite{Rahaman}. The authors arrived at a very interesting conclusion that flatness of the rotational
curves can be explained due to the presence of quark matter in halos (because of additional attraction from such matter). In our paper, we spread this quark
matter over the whole Universe and demonstrate that it can result in the late time acceleration of the Universe. Analogously, the authors of \cite{Rahaman}
also claimed that quark-gluon plasma on the global level behaves like dark energy.

The paper is structured as follows. In Sec. 2, we briefly remind the quark bag equations of state. In Sec. 3 and 4, we consider the influence of nuggets and
QGP-like perfect fluid on the late-time expansion of the Universe. The main results are briefly summarized in concluding Sec. 5.

\section{Equations of state in~the~quark-gluon bag model}

We first briefly remind the quark bag equation of state, a simple model of quark confinement. For vanishing chemical potential, $\mu=0,$ it is a system of two
equations

%%%%%%%%%%%%%%%%%%%%
\be{2.1a}
p_q(T) = A_q T^4 - B\ ,
\ee

\be{2.2a}
p_h(T) = A_h T^4\ .
\ee
%%%%%%
The first line corresponds to the "hot"\, phase of deconfined quarks and gluons, and the second one relates to confined particles, i.e. hadrons. A system of
strongly interacting particles, made of free quarks and gluons, is cooling down and meets the "cold"\, phase transforming in colorless hadrons. The
coefficients are defined by the degrees of freedom and are equal to:  $A_q\approx 1.75,\ \ A_h\approx 0.33$, $B=(A_q-A_h)T^4_c$ and $T_c\approx 200$ MeV.

Knowing the pressure, $p(T)$, for $\mu=0,$ one can easily calculate the remaining thermodynamical quantities, e.g., for the energy density we have
%%%%%%
\be{2.3}
\varepsilon(T)=T\frac{dp}{dT}-p\, .
\ee
%%%%%%

The above EoS is not unique. There is a number of interesting modifications \cite{LasloYad1,LasloYad2,Kallmann,LasloEch,Begun2004,Pisarski2006,Pisarski2007}.
First such modification was considered by C. K\"allmann \cite{Kallmann}, who introduced a temperature-dependent bag "constant", namely, by replacing in the
first line of the EoS, Eq. \rf{2.1a}, $B\rightarrow B(T)=\tilde BT$, where $\tilde B =(A_q-A_h)T^3_c$. This modification has immediate consequences, namely, by
producing a minimum in the "hot"\, line of the EoS, corresponding to metastable deeply supercooled states of the deconfined strongly interacting matter. Also,
it drives inflation of the Universe, as shown in \cite{LasloYad1,LasloYad2}. A detailed discussion of the above EoS and their consequences, both for the heavy
ion collisions and the early Universe, can be found in the review paper \cite{LasloEch}.

Since the idea of the present paper is that a small fraction of deconfined quarks and gluons survives to present days, we shall be interested in the "hot"\,
branch of the bag EoS. As we mentioned above, there is a number of different modifications of Eq. \rf{2.1a}. For our present purposes, however, two simple
representatives will be sufficient. They are the K\"allmann modified model (which we call Model I):
%%%%%%%%
\be{2.1}
p_q(T) = A_q T^4 - \tilde B T\equiv \bar A_1 T + \bar A_4 T^4\, ,
\ee
%%%%%%%%
and the original model (Model II) described by Eq. \rf{2.1a}:
%%%%%%
\be{2.2}
p_q(T) = A_q T^4 - B\equiv \bar A_0 + \bar A_4 T^4\, .
\ee
%%%%%%
It is worth noting that in these equations, we measure temperature in energetic units, i.e. in erg or MeV ($1 \mbox{MeV}\approx 0.1602\times 10^{-5}$ erg).
Then, pressure is measured in $\mbox{erg}^4$ or $\mbox{MeV}^4$ {\footnote{\label{dimension}Usually, the dimension of pressure is $\mbox{erg}/\mbox{cm}^3$. It
is not difficult to get the relation $1 \mbox{MeV}^4 \approx 2.09 \times 10^{26} \mbox{erg}/\mbox{cm}^3$. However, to transform to the usual units, it is more
convenient to redefine the coefficients as follows: $\bar A_i \to \widetilde A_i =\bar A_i /[(M_{Pl}c^2)^3 L_{Pl}^3]\, ,i=0,1,4$, where $M_{Pl}\approx
2.177\times 10^{-5}$g is the Planck mass and $L_{Pl}\approx1.616\times 10^{-33}$cm is the Planck length.}}.

%%%%%%%%%%%%%%%%%%%%%%%%%%%%%%%%%%%%%%%%%%%%%%%%%%%%%%%%%%%%%%%%%%%%%%%%%%%%%%%%%%%%%%%%%%%%%%%%%%%%%%%
%%%%%%%%%%%%%%%%%%%%%%%%%%%%%%%%%%%%%%%%%%%%%%%%%%%%%%%%%%%%%%%%%%%%%%%%%%%%%%%%%%%%%%%%%%%%%%%%%%%%%%%

\section{Quark nuggets}

As we wrote above, there is a possibility that after a phase transition from quark gluon plasma (QGP) to hadronic matter, a part of QGP was preserved in the
form of quark gluon nuggets \cite{Witten,Applegate,Farhi,Chandra}. They are isolated "islands" of QGP in a sea of a new hadronic phase. Now, we want to
investigate cosmological consequences of this assumption. Obviously, for considered models, a cosmological scenario strongly depends on thermodynamical
properties of QGP. We focus on two possible Eqs. \rf{2.1} and \rf{2.2}. With the help of standard thermodynamical Eq. \rf{2.3} we get the expressions for the
energy density:
%%%%%%
\be{2.4}
\varepsilon =3\bar A_4 T^4
\ee
%%%%%%
and
%%%%%%
\be{2.5}
\varepsilon =-\bar A_0 +3\bar A_4 T^4
\ee
%%%%%%
for Model I and Model II, respectively. Eqs. \rf{2.1}, \rf{2.2}, \rf{2.4} and \rf{2.5} describe the pressure and energy density inside of the nuggets. The
total pressure and energy density of all nuggets in the Universe can be calculated as follows. Let us take, e.g., Model I with Eq. \rf{2.1}. Then, for total
pressure of nuggets we get
%%%%%%
\be{2.6} P=\frac{\sum_i p_{qi} v_i}{V}= \frac{A_1 T +A_4 T^4}{a^3}\, , \ee where $p_{qi}$ is the pressure of the i-th nugget with the volume $v_i$ and
$V\propto a^3$ is the total volume of the Universe ($a$ is the scale factor of the Friedmann-Robertson-Walker metric). We consider the case where all nuggets
have the same pressure \rf{2.1} and their volumes are either constant or only slightly varying with time. The total volume of nuggets $\sum_i v_i$ is included
in the coefficients $A_1$ and $A_4$ (i.e. $A_{1,4}$ have dimension $\bar A_{1,4}\times \mbox{cm}^3$, so, taking into account the footnote \ref{dimension},
$A_1$ is dimensionless and $A_4$ has the dimension $\mbox{erg}^{-3}$). Therefore,
%%%%%
\be{2.7}
\frac{A_1}{A_4}=\frac{\bar A_1}{\bar A_4}= -0.8114 \; T_c^3\, .
\ee
%%%%%
Similarly, from Eq. \rf{2.4}, for the energy density of all nuggets we get:
%%%%%%
\be{2.8}
\mathcal{E} =\frac{3 A_4 T^4}{a^3}\, .
\ee
%%%%%
The same procedure holds for the Model II. Let us consider two models separately.

\subsection{Model I}

Here, the pressure and energy density of all nuggets are given by the above formulae \rf{2.6} and \rf{2.8}, respectively. In these formulae, temperature is a
function of the scale factor $a$: $T=T(a)$. Let us specify this dependence. From the energy conservation equation
%%%%%%
\be{2.9}
d(\mathcal{E} a^3) + Pd(a^3)=0
\ee
%%%%%%%
we can easily get
%%%%%
\be{2.10}
T=\left(\frac{\left(C/a\right)^{3/4}-A_1}{A_4}\right)^{1/3}\, .
\ee
%%%%%%
As we mentioned above, we consider the model where the coefficients $A_1 <0$ and $A_4>0$. In Eq. \rf{2.10}, $C\geq 0$ is the constant of integration which is
defined by the temperature $T_0$ and scale factor $a_0$ at the present time:
%%%%%%
\be{2.11}
C=\left(A_1+A_4T_0^3\right)^{4/3}a_0 = A_4^{4/3}\left(-0.8114\, T_c^3 +T_0^3\right)^{4/3}a_0\, .
\ee
%%%%%%
The temperature $T$ tends to the constant value when the scale factor approaches infinity:
%%%%%
\be{2.12} T \longrightarrow T_{\infty} = \left(\frac{-A_1}{A_4}\right)^{1/3} = 0.9327\, T_c\,  \quad \mbox{for} \quad a \to \infty\, , \ee
%%%%%
and the pressure goes asymptotically to zero: $P \to 0$. On the other hand, for $C\equiv 0$, the temperature is constant all the time $T \equiv T_{\infty}$,
and nuggets behave as a matter with zero pressure $P=0$. It is worth noting that in this model the temperature of the QNs at present time is not
arbitrary low, rather it is close to the critical temperature $T_c$ of the phase transition (see also Eq. \rf{2.32} for the Model II below).

We consider our Universe starting from the moment when we can drop the radiation. It is well known that the radiation dominated (RD) stage is much shorter than
the matter dominated (MD) stage. Hence, the neglect of the RD stage does not affect much the estimate of the lifetime of the Universe. Starting from the MD
stage, the first Friedmann equation for our model reads
%%%%%
\be{2.13}
3\frac{\mathcal{H}^2+K}{a^2}=\kappa \mathcal{E}+\kappa \varepsilon_0^{\mathrm{mat}}\left(\frac{a_0}{a}\right)^3+\Lambda\, ,
\ee
%%%%%
where we take into account the cosmological constant $\Lambda$ and the (usual + dark) matter with  the present value of the energy density
$\varepsilon_0^{\mathrm{mat}}$. In \rf{2.13}, $\mathcal{H}=a'/a=(da/d\eta)/a$, $\kappa=8\pi G_N/c^4$, $G_N$ is the gravitational constant and $K=\pm 1,0$ is
the spatial curvature. The conformal time $\eta$ is connected with the synchronous time $t$: $a d\eta = c dt$. Taking into account Eqs. \rf{2.8} and \rf{2.10},
we get for the Hubble parameter $H=(1/a)da/dt=(c/a^2) da/d\eta$ the following expression:
%%%%%%
\ba{2.14} H^2 = H_0^2\left\{\left[\beta\left(\frac{a_0}{a}\right)^3+\gamma\left(\frac{a_0}{a}\right)^\frac{9}{4}\right]^\frac{4}{3}\right.
+\left.\Omega_M\left(\frac{a_0}{a}\right)^3+\Omega_\Lambda+\Omega_K\left(\frac{a_0}{a}\right)^2\right\}\, , \ea
%%%%%%
where the cosmological parameters are
%%%%%
\ba{2.15} \Omega_M=\frac{c^2}{3H_0^2}\kappa\varepsilon_0^{\mathrm{mat}},\quad \Omega_\Lambda=\frac{c^2}{3H_0^2}\Lambda,\quad
\Omega_K=-K\left(\frac{c}{a_0H_0}\right)^2 \ea
%%%%%%
and we introduce the dimensionless parameters
%%%%%
\ba{2.16} \beta= \left(\frac{C}{a_0}\right)^{3/4}\frac{1}{A_4^{1/4}}\left(\frac{\kappa c^2}{a_0^3 H_0^2}\right)^{3/4},\quad  \gamma = -
\frac{A_1}{A_4^{1/4}}\left(\frac{\kappa c^2}{a_0^3 H_0^2}\right)^{3/4}\, . \ea
%%%%%
{}From the second Friedmann equation
%%%%%
\be{2.17}
\frac{2\mathcal{H}'+\mathcal{H}^2+K}{a^2}=-\kappa P+\Lambda
\ee
%%%%%
after some obvious algebra we obtain the deceleration parameter
%%%%%
\ba{2.18} -q&=&\frac{1}{aH^2}\frac{d^2a}{dt^2}=\left(\frac{H_0}{H}\right)^2\left\{\frac{\gamma}{2}\left[\beta\left(\frac{a_0}{a}\right)^{39/4}
+\gamma\left(\frac{a_0}{a}\right)^9\right]^{1/3}\right.\nn \\
&-&\left[\beta\left(\frac{a_0}{a}\right)^3+\gamma\left(\frac{a_0}{a}\right)^{9/4}\right]^{4/3}-
\left.\frac{\Omega_M}{2}\left(\frac{a_0}{a}\right)^3+\Omega_\Lambda\right\}\, . \ea
%%%%%
At the present time $t_0$, Eqs. \rf{2.14} and \rf{2.18} read
%%%%%%
\ba{2.19}
1&=&\left(\beta +\gamma\right)^{4/3}+
\Omega_M +\Omega_\Lambda+\Omega_K\, ,\\
\label{2.20}
-q_0&=&\frac{\gamma}{2}\left(\beta+\gamma\right)^{1/3}-\left(\beta+\gamma\right)^{4/3}- \frac{\Omega_M}{2}+\Omega_\Lambda\, .
\ea
%%%%%%
Additionally, we obtain from \rf{2.14} the differential equation
%%%%%
\be{2.21} d\widetilde{t}=\frac{\widetilde{a}d\widetilde{a}}{\sqrt{\left(\beta+\gamma\tilde{a}^{3/4}\right)^{4/3}+
\Omega_M\widetilde{a}+\Omega_\Lambda\widetilde{a}^4+\Omega_K\widetilde{a}^2}}\, , \ee
%%%%%
where we introduce the dimensionless quantities
%%%%%
\be{2.22}
\widetilde{a}=\frac{a}{a_0}, \quad \widetilde{t}=H_0t\, .
\ee
%%%%%
Therefore, the age of the Universe $\widetilde{t}_0$ is defined by the equality
%%%%%%
\be{2.23}
-\widetilde{t}_0=\int\limits_1^0\frac{\widetilde{a}d\widetilde{a}}{\sqrt{\left(\beta+\gamma\tilde{a}^{3/4}\right)^{4/3}+
\Omega_M\widetilde{a}+\Omega_\Lambda\widetilde{a}^4+\Omega_K\widetilde{a}^2}}\, .
\ee
%%%%%%

Now, we consider the case of the flat space $K=0 \to \Omega_K=0$. Then, Eq. \rf{2.19} reads
%%%%%
\be{2.24}
1=\left(\beta+\gamma\right)^{4/3}+ \Omega_M+\Omega_\Lambda\, .
\ee
%%%%%
Eq. \rf{2.20} demonstrates that accelerated expansion of the Universe at the present time (i.e. $-q_0 >0$) can be ensured by the first and the last terms on
the right side of this equation. It is tempting to explain the acceleration only at the expense of the first term, i.e. due to the presence of QNs
when the cosmological constant is absent. However, simple analysis of Eqs. \rf{2.20} and \rf{2.24} in the case $\Omega_{\Lambda}=0$ shows that the acceleration
$-q_0>0$ is achieved only for $\beta <0$ that contradicts our model. The inclusion of the negative curvature $\Omega_K >0$ does not affect this conclusion due
to the smallness of $\Omega_K$.

Nevertheless, QNs can contribute to the dark matter if they weekly interact with usual baryon matter and light, or they can explain the problem of
missing baryons \cite{Bregman,Gupta} if their interaction with usual matter is not negligible. As we have mentioned above, nuggets behave as matter either
asymptotically when $a\to\infty$ or for all time in the case $C=0 \to \beta =0$. In the latter case we can exactly restore the $\Lambda$CDM model so long as
Eqs. \rf{2.20} and \rf{2.24} take the usual form for this model:
%%%%%%
\be{2.25}
1=\Omega_{M,total} +\Omega_{\Lambda}
\ee
%%%%%
and
%%%%%%
\be{2.26}
-q_0 = -\frac12 \Omega_{M,total}+ \Omega_{\Lambda} \quad \Rightarrow \quad \Omega_\Lambda=\frac{1}{3}-\frac{2}{3}q_0\, ,
\ee
%%%%%%
where $\Omega_{M,total} \equiv \gamma^{4/3}+\Omega_M$. Let $\Omega_M$ correspond to just the visible matter. According to observations, $\Omega_M \approx
0.04$. Then, we can easily restore the parameters of the $\Lambda$CDM model. For example, taking the deceleration parameter $q_0 \approx -0.595$, as in the
$\Lambda$CDM model \cite{WMAP7,WMAP9}, we get $\Omega_\Lambda\approx 0.73$ and $\gamma\approx 0.33 \to \gamma^{4/3} \approx 0.23$. Therefore, $\Omega_{M,total}
\approx 0.27$. For the age of the Universe, we get from \rf{2.21} (where we should put $\beta=0$, $\Omega_K=0$) $\widetilde t_0 \approx 1 \Rightarrow t_0
\approx H_0^{-1} \sim 13.7\times 10^9$yr. Hence, weekly interacting QNs may be candidates for dark matter.
%%%%%%%%%%%%%%%%%%%%%%%%%%%%%%%%%%%%%%%%%%%%%%%%%%%%%%%%%%%
%%%%%%%%%%%%%%%%%%%%%%%%%%%%%%%%%%%%%%%%%%%%%%%%%%%%%%%%%%%%

\subsection{Model II}

Quark nuggets for the Model II are defined by the thermodynamical functions \rf{2.2} and \rf{2.5}. Similar to Eqs. \rf{2.6} and \rf{2.8}, the total pressure
and energy density of all nuggets in the Universe are
%%%%
\ba{2.27}
P&=&\frac{A_0+A_4T^4}{a^3}\, ,\\
\label{2.28}\mathcal{E}&=&\frac{-A_0+3A_4T^4}{a^3}\, ,
\ea
%%%%%%
where
%%%%%
\be{2.29}
\frac{A_0}{A_4}=\frac{\bar A_0}{\bar A_4}= -0.8114 \; T_c^4\, .
\ee
%%%%%
In this model, $A_0<0$ and $A_4>0$. Taking into account the footnote \ref{dimension}, we may conclude that the coefficients $A_0$ and $A_4$ have dimensions erg
and $\mbox{erg}^{-3}$, respectively. For the thermodynamical functions \rf{2.27} and \rf{2.28}, energy conservation Eq. \rf{2.9} gives the following dependence
of the temperature on the scale factor:
%%%%%%
\be{2.30}
T=\left(\frac{(\widetilde C/a)-A_0}{A_4}\right)^{1/4}\, ,
\ee
%%%%%%
where $\widetilde C\geq 0$ is the constant of integration which is defined by the temperature $T_0$ and the scale factor $a_0$ at the present time:
%%%%%%
\be{2.31}
\widetilde C=\left(A_0+A_4T_0^4\right)a_0 = A_4\left(-0.8114\, T_c^4 +T_0^4\right)a_0\, .
\ee
%%%%%%
Similar to the previous case, the temperature $T$ tends to the constant value when the scale factor approaches infinity:
%%%%%
\be{2.32} T \longrightarrow T_{\infty} = \left(\frac{-A_0}{A_4}\right)^{1/4} = 0.9491\, T_c \quad \mbox{for} \quad a \to \infty\, , \ee
%%%%%
and the pressure goes asymptotically to zero: $P \to 0$. On the other hand, for $\widetilde C\equiv 0$, the temperature is constant all the time $T \equiv
T_{\infty}$, and QNs behave as a matter with zero pressure $P=0$.

In this model, the pressure and energy density have simple dependence on the scale factor:
%%%%%
\ba{2.33}
P(a)&=&\frac{\widetilde C}{a^4}\, ,\\
\label{2.34} \mathcal{E}(a)&=&3\frac{\widetilde C}{a^4}-4\frac{A_0}{a^3}\, .
\ea
%%%%%
Formally, such perfect fluid can be considered as a mixture of radiation and matter. However, for ordinary radiation $T\sim 1/a$.

{}From the first Friedmann equation \rf{2.13}, we obtain the expression for the Hubble parameter:
%%%%%
\ba{2.35} H^2 = H_0^2\left\{\beta\left(\frac{a_0}{a}\right)^4+\gamma\left(\frac{a_0}{a}\right)^3\right. +
\left.\Omega_M\left(\frac{a_0}{a}\right)^3+\Omega_\Lambda+\Omega_K\left(\frac{a_0}{a}\right)^2\right\}\, , \ea
%%%%%
where the cosmological parameters are defined in \rf{2.15} and the dimensionless parameters $\beta$ and $\gamma$ are
%%%%%
\be{2.36}
\beta = \frac{\widetilde C}{a_0}\left(\frac{\kappa c^2}{a_0^3 H_0^2}\right)\, ,\quad \gamma = -\frac{4 A_0}{3}\left(\frac{\kappa c^2}{a_0^3 H_0^2}\right)\, .
\ee
%%%%%
Therefore, the age of the Universe $\widetilde{t}_0$ is defined by the equality
%%%%%
\be{2.37}
-\widetilde{t}_0=\int\limits_1^0\frac{\widetilde{a}d\widetilde{a}}{\sqrt{\beta+\gamma\tilde{a}+
\Omega_M\widetilde{a}+\Omega_\Lambda\widetilde{a}^4+\Omega_K\widetilde{a}^2}}\, .
\ee
%%%%%%
The second Friedmann equation \rf{2.17} results in the deceleration parameter
%%%%%
\ba{2.38} -q=\frac{1}{aH^2}\frac{d^2a}{dt^2}=\left(\frac{H_0}{H}\right)^2\left\{
-\beta\left(\frac{a_0}{a}\right)^4-\frac{1}{2}\gamma\left(\frac{a_0}{a}\right)^3\right.-\left.\frac{1}{2}\Omega_M\left(\frac{a_0}{a}\right)^3+
\Omega_\Lambda\right\}\, . \ea
%%%%%
At the present time $t_0$, Eqs. \rf{2.35} and \rf{2.38} read
%%%%%
\ba{2.39}
1&=&\beta+\gamma+\Omega_M+\Omega_\Lambda+\Omega_K\, ,\\
\label{2.40}-q_0&=&-\beta-\frac{1}{2}\gamma+\Omega_\Lambda-\frac{1}{2}\Omega_M\, .
\ea
%%%%%

It can be easily seen that similar to the previous model we also reproduce the standard $\Lambda$CDM model in the case of the flat space $\Omega_K=0$ and
$\widetilde C=0 \to \beta =0$. The only difference is that in Eqs. \rf{2.25} and \rf{2.26} $\Omega_{M,total} = \gamma +\Omega_M$. For example, if we take
$\Omega_M\approx 0.04$ and $q_0\approx -0.595$, then we get $\Omega_\Lambda\approx 0.73$ and $\gamma\approx 0.23 \to \Omega_{M,total} \approx 0.27$ as in the
$\Lambda$CDM model. For these parameters, the age of the Universe is $\widetilde t_0 \approx 1 \Rightarrow t_0 \approx H_0^{-1} \sim 13.7\times 10^9$yr.
Hence,we again arrive at the conclusion that weekly interacting QNs may be candidates for dark matter.
%%%%%%%%%%%%%%%%%%%%%%%%%%%%%%%%%%%%%%%%%%%%%%%%%%%%%%%%%%%%%%%%%%%%%%%%%%%%%%%%%%5
%%%%%%%%%%%%%%%%%%%%%%%%%%%%%%%%%%%%%%%%%%%%%%%%%%%%%%%%%%%%%%%%%%%%%%%%%%%%%%%%%%%%
%%%%%%%%%%%%%%%%%%%%%%%%%%%%%%%%%%%%%%%%%%%%%%%%%%%%%%%%%%%%%%%%%%%%%%%%%%%%%%%%%%%%

\section{QGP-like perfect fluid}

Above, we considered a scenario where QNs form after the phase transition the "isolated islands" in the sea of baryon matter. In this section we
suppose a less physically motivated model where a part of quark gluon plasma survived after the phase transition in the form of the homogeneously and
isotropically distributed perfect fluid. This perfect fluid has the thermodynamical functions of the form \rf{2.1}, \rf{2.4} or \rf{2.2}, \rf{2.5}. We do not
know what part of QGP survived (see however some estimate at the very end of this section). So, we demand only that the ratio between coefficients $\bar A_i$
was preserved. Maybe, it is more correct to speak about some unknown perfect fluid with the thermodynamical functions motivated by the QGP. Therefore, we call
this fluid as a QGP-like perfect fluid. We are going to investigate the cosmological consequences of such proposal.

In general, we can consider the pressure of the form
%%%%%%
\be{a.1}
p(T)=\widehat A_0+ \widehat A_1T+ \widehat A_2T^2+ \widehat A_3T^3+ \widehat A_4T^4\, ,
\ee
%%%%%%
which, via Eq. \rf{2.3}, results in the energy density
%%%%%%
\be{a.2}
\varepsilon(T)=-\widehat A_0+ \widehat A_2T^2+2 \widehat A_3T^3+3 \widehat A_4T^4\, .
\ee
%%%%%%
Eq. \rf{2.9} leads to the differential equation
%%%%%
\be{a.3} \frac{da}{a}=-\frac{\left(2\widehat A_2+6\widehat A_3T+12\widehat A_4T^2\right)}{3\left(\widehat A_1+2\widehat A_2T+3\widehat A_3T^2+4\widehat
A_4T^3\right)}dT\, . \ee
%%%%%
The solution of this equation enables to determine the dependence $T=T(a)$. Unfortunately, there is no solution of \rf{a.3} in elementary functions. Therefore,
we consider two particular models by analogy with the previous section.

\subsection{Model I}

First, we consider the case $\widehat A_0=0$, $\widehat A_2=0$, $\widehat A_3=0$. As we mentioned above, the coefficients $\widehat A_1$ and $\widehat A_4$
satisfy the condition similar to \rf{2.7}:
%%%%%
\be{a.4}
\frac{\widehat A_1}{\widehat A_4}=\frac{\bar A_1}{\bar A_4}= -0.8114 \; T_c^3\, .
\ee
%%%%
Therefore, we consider the case $\widehat A_1 <0$ and $\widehat A_4 >0$. Following the footnote \ref{dimension}, the coefficients $\widehat A_1$ and $\widehat
A_4$ have dimensions $\mbox{cm}^{-3}$ and $\mbox{erg}^{-3}\mbox{cm}^{-3}$, respectively.

Integrating \rf{a.3}, we get
%%%%%%
\ba{a.5} T=\left(\frac{\widehat C^3-\widehat A_1a^3}{4\widehat A_4a^3}\right)^{1/3}\quad \Rightarrow \quad \varepsilon(T)=3\widehat A_4\left(\frac{\widehat
C^3-\widehat A_1a^3}{4\widehat A_4a^3}\right)^{4/3}\, , \ea
%%%%%%
where $\widehat C\geq 0$ is the dimensionless constant of integration which is defined by the temperature $T_0$ and the scale factor $a_0$ at the present time:
%%%%%%
\ba{a.6} \widehat C=\left(\widehat A_1+4\widehat A_4T_0^3\right)^{1/3}a_0= \widehat A_4^{1/3}\left(-0.8114\, T_c^3 +4 T_0^3\right)^{1/3}a_0\, . \ea
%%%%%%
The difference between the first equation in \rf{a.5} and Eq. \rf{2.10} is due to the prefactor $1/a^3$ in \rf{2.6} and \rf{2.8}. The temperature $T$ tends to
the constant value when the scale factor approaches infinity
%%%%%
\be{a.7} T \longrightarrow T_{\infty} = \left(\frac{-\widehat A_1}{4\widehat A_4}\right)^{1/3} = 0.5876\, T_c \quad \mbox{for} \quad a \to \infty\, . \ee
%%%%%
It can be easily verified that, in the limit $a \to \infty$, the energy density $\varepsilon \to (3/4)[\widehat A_1^4/(4\widehat A_4)]^{1/3}$ and the pressure
$p \to - (3/4)[\widehat A_1^4/(4\widehat A_4)]^{1/3}$, i.e. the perfect fluid has asymptotically the vacuum-like equation of state $p=-\varepsilon$. If
$\widehat C \equiv 0$, then the perfect fluid has this equation of state for all time.

{}From the first Friedmann equation \rf{2.13}, we get the Hubble parameter
%%%%%
\ba{a.8} H^2=H_0^2\left\{\left[\beta \left(\frac{a_0}{a}\right)^3+\gamma
\right]^{4/3}\right.+\left.\Omega_M\left(\frac{a_0}{a}\right)^3+\Omega_\Lambda+\Omega_K\left(\frac{a_0}{a}\right)^2\right\}\, , \ea
%%%%%
where
%%%%%
\ba{a.9} \beta = \frac{\widehat C^3}{4(a_0^3 \widehat A_4)^{1/4}}\left(\frac{\kappa c^2}{a_0^3 H_0^2}\right)^{3/4},\quad \gamma = - \frac{\widehat A_1
a_0^3}{4(a_0^3 \widehat A_4)^{1/4}} \left(\frac{\kappa c^2}{a_0^3 H_0^2}\right)^{3/4}\, . \ea
%%%%%
We use Eq. \rf{a.8} to find the age of the Universe:
%%%%%
\be{a.10} -\widetilde{t}_0=\int\limits_1^0\frac{\widetilde{a} d\widetilde{a}}{\sqrt{\left(\beta+\gamma\widetilde{a}^3
\right)^{4/3}+\Omega_M\widetilde{a}+\Omega_{\Lambda}\widetilde{a}^4+\Omega_K\widetilde{a}^2}}\, . \ee
%%%%%
{}From the second Friedmann equation \rf{2.17}, the deceleration parameter is
%%%%%
\ba{a.11} -q=\left(\frac{H_0}{H}\right)^2\left\{2\gamma \left[\beta\left(\frac{a_0}{a}\right)^3+\gamma\right]^{1/3}\right. -
\left.\left[\beta\left(\frac{a_0}{a}\right)^3+\gamma\right]^{4/3}-\frac{1}{2}\Omega_M\left(\frac{a_0}{a}\right)^3 +\Omega_\Lambda \right\}\, . \ea
%%%%%
At the present time $t_0$, Eqs. \rf{a.8} and \rf{a.10} read
%%%%%
\ba{a.12}
1&=&\left(\beta +\gamma \right)^{4/3}+\Omega_M+\Omega_\Lambda+\Omega_K\, ,\\
\label{a.13}-q_0&=&2\gamma \left(\beta+\gamma\right)^{1/3}-\left(\beta+\gamma\right)^{4/3}-\frac{1}{2}\Omega_M+\Omega_\Lambda\, .
\ea
%%%%%

The latter equation indicates that the acceleration of the Universe (i.e. $-q_0>0$) in this model can originate from the first and the last terms on the
right-hand side of this equation. Up to now, the nature of the cosmological constant is still unclear. So, we try to explain the acceleration without it, i.e.
we suppose that $\Omega_{\Lambda}=0$. It is also well known that our Universe is very flat \cite{WMAP7,WMAP9}. Hence, we put $\Omega_K=0$.

Let us consider two particular cases. The first one corresponds to the choice $\beta=0$. In this case, Eqs. \rf{a.12} and \rf{a.13} take the form
%%%%%
\ba{a.14}
1&=&\Omega_M + \Omega_{\Lambda,qgp}\, ,\\
\label{a.15}-q_0&=&-\frac12\Omega_M + \Omega_{\Lambda,qgp}\, ,
\ea
%%%%%%
where $\Omega_{\Lambda,qgp}\equiv \gamma^{4/3}$. These equations reproduce exactly the standard $\Lambda$CDM model. Therefore, if we take $-q_0 \approx 0.595$
and $\Omega_M \approx 0.27$, then we get $\Omega_{\Lambda,qgp}\approx 0.73$ and for the age of the Universe $\widetilde t_0 \approx 1 \Rightarrow t_0 \approx
H_0^{-1} \sim 13.7\times 10^9$yr.

The second case with $\beta \neq 0$ is a bit more complicated, but also more interesting. Because the nature of dark matter is unclear and, according to the
observations \cite{WMAP7,WMAP9}, the visible matter has $\Omega_M\approx0.04$, we shall take this value as a total contribution of matter. The other
experimental restriction follows from the age of globular clusters which is $11\div16$ Gyr \cite{Krauss}. Therefore, the Universe cannot be younger. Now, we
shall demonstrate that in our model we can satisfy this limitation if we suppose only $\Omega_M \approx0.04$ and $-q_0 \approx 0.595$ (as in the $\Lambda$CDM
model).

{}From Eqs. \rf{a.12} and \rf{a.13} (where $\Omega_{\Lambda}=\Omega_K=0$), we express the parameters $\beta$ and $\gamma$ via $\Omega_M$ and $q_0$:
%%%%%%
\be{a.16}
\beta = \frac{2-3\Omega_M+2q_0}{4(1-\Omega_M)^{1/4}}\, ,\quad
\gamma = \frac{2-\Omega_M-2q_0}{4(1-\Omega_M)^{1/4}}\, .
\ee
%%%%%%
Then, for $\Omega_M \approx 0.04$ and $-q_0 \approx 0.595$, we get $\beta\approx 0.174$ and $\gamma\approx 0.796$. For these values of $\beta$ and $\gamma$,
the age of the Universe is $\widetilde t_0\approx0.892 \Rightarrow  t\approx 12.2$ Gyr, in agreement with the experimental data. Roughly speaking, the
parameter $\gamma$ is responsible for the accelerated expansion of the Universe, and the parameter $\beta$ plays the role of dark substance.

%%%%%%%%%%%%%%%%%%%%%%%%%%%%%%%%%%%%%%%%%%%%%%%%%%%%%%%%%%%%%%%%%%%%%%%%%%%%%%%%%%%%%%%%%%%%%%%%%%%%%%%%%%%%%%%%%%%%

\subsection{Model II}

Let us consider now the case $\widehat A_1=0$, $\widehat A_2=0$, $\widehat A_3=0$. The coefficients $\widehat A_0$ and $\widehat A_4$ satisfy the condition
similar to \rf{2.29}:
%%%%%
\be{a.17} \frac{\widehat A_0}{\widehat A_4}=\frac{\bar A_0}{\bar A_4}= -0.8114 \; T_c^4\, , \ee
%%%%
where $\widehat A_0 <0$ and $\widehat A_4 >0$, and they have dimensions $\mbox{erg}\times \mbox{cm}^{-3}$ and $\mbox{erg}^{-3}\mbox{cm}^{-3}$, respectively.

Integrating \rf{a.3}, we get
%%%%%%
\be{a.18}
T=\frac{\overline{C}}{a}\, ,
\ee
%%%%%%
where $\overline C\geq 0$ is the constant of integration which is defined by the temperature $T_0$ and the scale factor $a_0$ at the present time: $C=a_0 T_0$.
Therefore, the pressure and energy density depend on the scale factor as follows:
%%%%%%
\be{a.19}
\varepsilon =-\widehat A_0+3\widehat A_4\left(\frac{\overline C}{a}\right)^4,\quad p =\widehat A_0+\widehat A_4\left(\frac{\overline C}{a}\right)^4\, .
\ee
%%%%%%
Hence, such perfect fluid can be formally considered as a mixture of vacuum and radiation. This conclusion is also confirmed by the form of the Friedmann
equations for this model:
%%%%%%%
\ba{a.20} H^2&=&H_0^2\left[\beta\left(\frac{a_0}{a}\right)^4 +\gamma \right.
+\left.\Omega_M\left(\frac{a_0}{a}\right)^3+\Omega_\Lambda+\Omega_K\left(\frac{a_0}{a}\right)^2\right]\, ,\\
\label{a.21} -q&=&\left(\frac{H_0}{H}\right)^2\left[\gamma -\beta\left(\frac{a_0}{a}\right)^4\right.
-\left.\frac{1}{2}\Omega_M\left(\frac{a_0}{a}\right)^3+\Omega_\Lambda\right]\, , \ea
%%%%%%
where
%%%%%
\be{a.22} \beta = \frac{\widehat A_4 \overline{C}^4}{a_0}\left(\frac{\kappa c^2}{a_0^3 H_0^2}\right)\, ,\quad \gamma = -\frac{\widehat A_0 a_0^3}{3}
\left(\frac{\kappa c^2}{a_0^3 H_0^2}\right)\, . \ee
%%%%%
For the age of the Universe we have
%%%%%
\be{a.23}
-\widetilde{t}_0=\int\limits_1^0 \frac{\widetilde{a}d\widetilde{a}}{\sqrt{\beta+\gamma\widetilde{a}^4
+\Omega_M\widetilde{a}+\Omega_\Lambda\widetilde{a}^4 +\Omega_K\widetilde{a}^2}}\, .
\ee
%%%%
Obviously, the parameter $\gamma$ plays the role of the cosmological constant. So, we may omit $\Omega_{\Lambda}$ in above equations. According to the
observations, we may also put $\Omega_K=0$ because of its smallness. We restore exactly the $\Lambda$CDM model, e.g., with the choice $\beta =0$ (then,
$\gamma\approx 0.73$). Therefore, in this model the cosmological constant arises due to QGP. Let us estimate the fraction of QGP which should remain after the
phase transition to get the observable acceleration. It is clear that parameters $\bar A_0$ and $\widehat A_0$ play the role of the vacuum energy density
before and after the phase transition, respectively: $\varepsilon_{V,in}=-\bar A_0$, $\varepsilon_{V,fin}=-\widehat A_0$. The initial vacuum energy density
$\varepsilon_{V,in}=-\bar A_0 \approx 1.42\, T_c^4 \sim 2\times10^9\mbox{MeV}^4$ \cite{LasloEch}. For $\gamma \approx 0.73$ from \rf{a.22} we get
$\varepsilon_{V,fin}=-\widehat A_0\sim 6\times10^{-9}\mbox{erg}\times\mbox{cm}^{-3}\approx 3\times 10^{-35}\mbox{MeV}^4$. Thus,
$\varepsilon_{V,fin}/\varepsilon_{V,in}\sim 10^{-44}$.

\section{Conclusions}

Our paper was devoted to two great challenges of modern cosmology and high energy physics dubbed dark matter and dark energy. Up to now, there is no
satisfactory explanation for both of them. In our paper, we proposed a possible solution to these problems. For this purpose, we considered the expansion of
the present Universe, using the "hot"\,, i.e. the quark-gluon branch of the bag EoS. Although we made reference to the role of this type of the EoS during the
early universe, namely its inflation phase, here we postponed possible speculations about the continuous evolution of the universe, within the present
formalism, from it early, quark-gluon stage to the present days, admitting only the possible continuity in the existence in the present Universe of a small
fraction of colored objects -- quarks and gluons -- which escaped hadronization. We considered two different scenarios.

In the first scenario, we supposed that the colored objects survived in the form of isolated islands, called quark-gluon nuggets, in a sea of a hadronic phase.
In the second scenario, we assumed that a very small fraction of colored objects -- quarks and gluons -- survived after the phase transition in the form of a
perfect fluid, called the QGP-like perfect fluid, uniformly spread within the colorless hadronic medium. Obviously, these cosmological scenarios are defined by
EoS of quark gluon plasma (QGP). We focus on two possible Eqs. \rf{2.1} and \rf{2.2} dubbed Model I and Model II, respectively. We have shown that within
considered scenarios, there are no fundamental differences in the obtained conclusions for these models. For the nugget-scenario, we have shown that weekly
interacting (with visible matter) QNs can play the role of dark matter for both of the models. In the case of QGP-like fluid, we have demonstrated
that this fluid can play the role of dark energy  providing the late-time accelerating expansion of the Universe for both of the models. Moreover, we defined
that, to be in agreement with observations, only $10^{-44}$ part of the colored objects should survive after the phase transition. Therefore, the considered
scenarios provide new possible ways of solving the problems of dark matter and dark energy.

%%%%%%%%%%%%%%%%%%%%%%%%%%%%%%%%%%%%%%%%%%%%%%%%%%%%%%%%%%%%%%%%%%%%%%%%%%%%%%%%
%%%%%%%%%%%%%%%%%%%%%%%%%%%%%%%%%%%%%%%%%%%%%%%%%%%%%%%%%%%%%%%%%%%%%%%%%%%%%%%%

\section*{Acknowledgements}

This work was supported by the "Cosmomicrophysics-2" programme of the Physics and Astronomy Division of the National Academy of Sciences of Ukraine. The work
of M. Eingorn was supported by NSF CREST award HRD-0833184 and NASA grant NNX09AV07A.

%%%%%%%%%%%%%%%%%%%%%%%%%%%%%%%%%%%%%%%%%%%%%%%%%%%%%%%%%%%%%%%%%%%%%%%%%%%%%%%%%%%%%%%
%%%%%%%%%%%%%%%%%%%%%%%%%%%%%%%%%%%%%%%%%%%%%%%%%%%%%%%%%%%%%%%%%%%%%%%%%%%%%%%%%%%%%%%


\begin{thebibliography}{99}
\bibitem{LasloYad1}
L.L. Jenkovszky, B. K\"ampfer and V.M. Sysoev, Z. Phys. C, Particles and Fields {\bf 48} (1990) 147.
%%%%%%%%
\bibitem{LasloYad2}
V.G. Boyko, L.L. Jenkovszky, B. K\"ampfer and V.M. Sysoev, J. Nucl. Phys. {\bf 51} (1990) 1134.
%%%%%%%
\bibitem{Tillmann} T. Boeckel and J. Schaffner-Bielich, Phys. Rev. D {\bf 85} (2012) 103506; arXiv:astro-ph/1105.0832.
%%%%%%%%%
\bibitem{Witten}
E. Witten, Phys. Rev. D {\bf 30} (1984) 272.
%%%%%%%
\bibitem{Applegate}
A. Applegate and C.J. Hogan, Phys. Rev. D {\bf 31} (1985) 3037.
%%%%%%%
\bibitem{Farhi}
E. Farhi and R.L. Jaffe, Phys. Rev. D {\bf 30} (1984) 2379.
%%%%%%%
\bibitem{Chandra}
D. Chandra and A. Goyal, Phys. Rev. D {\bf 62} (2000) 063505; arXiv:hep-ph/9903466.
%%%%%%%
%\bibitem{Tillmann} Tillmann Boeckel, Jurgen Schaffner-Bielich, {\it A little inflation at the cosmological QCD phase transition}, arXiv:1105.0832.
%%%%%%%%
\bibitem{Bhatta}
A. Bhattacharyya et al., Nucl. Phys. A {\bf 661} (1999) 629; arXiv:hep-ph/9907262.
%%%%%%%%
\bibitem{Bhatta2}
A. Bhattacharyya et al., Phys. Rev. D {\bf 61} (2000) 083509; arXiv:hep-ph/9901308.
%%%%%%%%
\bibitem{BASR}
P. Bhattacharjee, J. Alam, B. Sinha and S. Raha, Phys. Rev. D {\bf 48} (1993) 4630.
%%%%%%%%
\bibitem{ARS}
J. Alam, S. Raha and B. Sinha, Astrophys J. {\bf 513} (1999) 572; arXiv:astro-ph/9704226.
%%%%%%%%
\bibitem{Gh}
S. Ghosh, {\it Astrophysics of Strange Matter}, Plenary Talk at 2008 Quark Matter, Jaipur, India; arXiv:astro-ph/0807.0684.
%%%%%%%%%
\bibitem{Kalam}
M. Kalam et al., {\it A relativistic model for Strange Quark Star}; arXiv:gr-qc/1205.6795.
%%%%%%%%%
\bibitem{Zhitnitsky}
A.R. Zhitnitsky, JCAP {\bf 10} (2003) 010; arXiv:hep-ph/0202161.
%%%%%%%%%
\bibitem{Perez}
M.A. Perez-Garcia, J. Silk and J.R. Stone, Phys. Rev. Lett. {\bf 105} (2010) 141101; arXiv:astro-ph/1007.1421.
%%%%%%%
\bibitem{Drake}
J.J. Drake et al.,
Astrophys. J. {\bf 572} (2002) 996; arXiv:astro-ph/0204159.
%%%%%%%%%
\bibitem{Madsen}
J. Madsen, Lect. Notes Phys. {\bf 516} (1999) 162; arXiv:astro-ph/9809032.
%%%%%%%%%
\bibitem{Madsen2}
J. Madsen and J.M. Larsen, Phys. Rev. Lett. {\bf 90} (2003) 121102; arXiv:astro-ph/0211597.
%%%%%%%%
\bibitem{Madsen3}
J. Madsen, Phys. Rev. D {\bf 71} (2005) 014026; arXiv:astro-ph/0411538.
%%%%%%%
\bibitem{Rahaman}
F. Rahaman et al., Phys. Lett. B {\bf 714} (2012) 131; arXiv:gr-qc/1203.6649.
%%%%%%%%
\bibitem{Kallmann}
C.G. K\"allmann, Phys. Lett. B {\bf 134} (1984) 363.
%%%%%%%%
\bibitem{LasloEch}
V.G. Boyko, L.L. Jenkovszky and V.M. Sysoev, EChAYa {\bf 22} (1991) 675.
%%%%%%%%
\bibitem{Begun2004}
V.V. Begun, M.I. Gorenstein and O.A. Mogilevsky, Int. J. Mod. Phys. E {\bf 20} (2011) 1805; arXiv:hep-ph/1004.0953.
%%%%%%%%
\bibitem{Pisarski2006}
R.D. Pisarski, Phys. Rev. D {\bf 74} (2006) 121703.
%%%%%%%%
\bibitem{Pisarski2007}
R.D. Pisarski, Progr. Theor. Phys. Suppl. {\bf 168} (2007) 276.
%%%%%%%
\bibitem{Bregman}
J.N. Bregman, Ann. Rev. Astron. Astrophys. {\bf 45} (2007) 221; arXiv:astro-ph/0706.1787.
%%%%%%%
\bibitem{Gupta}
A. Gupta, S. Mathur, Y. Krongold, F. Nicastro and M. Galeazzi, The Astrophysical Journal Letters, {\bf 756} (2012) L8; arXiv:astro-ph/1205.5037.
%%%%%%%
\bibitem{WMAP7}
E. Komatsu et al, Astrophys. J. Suppl. {\bf 192} (2011) 18; arXiv:astro-ph/1001.4538.
%%%%%%%
\bibitem{WMAP9}
G. Hinshaw et al, Nine-Year Wilkinson Microwave Anisotropy Probe (WMAP) Observations: Cosmological Parameter Results (2012); arXiv:astro-ph/1212.5226.
%%%%%%%%%
\bibitem{Krauss}
L.M. Krauss and B. Chaboyer, Science, Volume 299, Issue 5603 (2003) 65-70; L. M. Krauss, The State of the Universe: Cosmological Parameters (2002);
arXiv:astro-ph/0301012.
\end{thebibliography}
\end{document}